\documentclass[aip, pre, twocolumn, superscriptaddress, reprint]{revtex4-2}
\usepackage[T1]{fontenc}
\usepackage{times}
\usepackage{graphicx}
\usepackage{blkarray}

\usepackage{amsfonts, amsmath,amssymb, bm, graphicx}
\usepackage{siunitx}

\usepackage[english]{babel}
\usepackage{lipsum}

\newcommand{\figref}[2]{\hyperref[#1]{\ref{#1}(#2)}}
\usepackage{physics}
\usepackage{lipsum}
\usepackage{changes}

\usepackage[colorlinks=true,allcolors=blue]{hyperref}
\usepackage{footmisc}

\usepackage{upgreek}

\usepackage{soul}

\begin{document}

{
\makeatletter
\def\frontmatter@thefootnote{%
\altaffilletter@sw{\@fnsymbol}{\@fnsymbol}{\csname c@\@mpfn\endcsname}%
}%

\makeatother
\title{Finite-element dynamic-matrix approach for propagating spin waves: Extension to mono- and multilayers of arbitrary spacing and thickness}

\author{L. K\"orber}\email{l.koerber@hzdr.de}
\affiliation{Helmholtz-Zentrum Dresden - Rossendorf, Institut f\"ur Ionenstrahlphysik und Materialforschung, D-01328 Dresden, Germany}
\affiliation{Fakultät Physik, Technische Universit\"at Dresden, D-01062 Dresden, Germany}

\author{A. Hempel}
\affiliation{Helmholtz-Zentrum Dresden - Rossendorf, Institut f\"ur Ionenstrahlphysik und Materialforschung, D-01328 Dresden, Germany}
\affiliation{Fakultät Physik, Technische Universit\"at Dresden, D-01062 Dresden, Germany}

\author{A. Otto}
\affiliation{Fakultät Physik, Technische Universit\"at Dresden, D-01062 Dresden, Germany}

\author{R. A. Gallardo}
\affiliation{Departamento de Física, Universidad T\'ecnica Federico Santa Mar\'ia, Avenida Espa\~{n}a 1680, Valpara\'iso, Chile}

\author{Y. Henry}
\affiliation{Institut de Physique et Chimie des Mat\'eriaux de Strasbourg, UMR 7504, CNRS and Universit\'e de Strasbourg, B.P. 43, F-67037 Strasbourg Cedex 2, France}

\author{J. Lindner}
\affiliation{Helmholtz-Zentrum Dresden - Rossendorf, Institut f\"ur Ionenstrahlphysik und Materialforschung, D-01328 Dresden, Germany}

\author{A. Kákay}
\affiliation{Helmholtz-Zentrum Dresden - Rossendorf, Institut f\"ur Ionenstrahlphysik und Materialforschung, D-01328 Dresden, Germany}

\date{\today}

\begin{abstract}

In our recent work (Ref.~\citenum{korberFiniteelementDynamicmatrixApproach2021}), we presented an efficient numerical method to compute dispersions and mode profiles of spin waves in waveguides with translationally invariant equilibrium magnetization. A finite-element method (FEM) allowed to model two-dimensional waveguide cross sections of arbitrary shape but only finite size. Here, we extend our FEM propagating-wave dynamic-matrix approach from finite waveguides to the important cases of infinitely-extended mono- and multilayers of arbitrary spacing and thickness. To obtain the mode profiles and frequencies, the linearized equation of motion of magnetization is solved as an eigenvalue problem on a one-dimensional line-trace mesh, defined along the normal direction of the layers. Being an important contribution in multilayer systems, we introduce interlayer exchange into our FEM approach. With the calculation of dipolar fields being the main focus, we also extend the previously presented plane-wave Fredkin-Koehler method to calculate the dipolar potential of spin waves in infinite layers. The major benefit of this method is that it avoids the discretization of any non-magnetic material like non-magnetic spacers in multilayers. Therefore, the computational effort becomes independent on the spacer thicknesses. Furthermore, it keeps the resulting eigenvalue problem sparse, which therefore, inherits a comparably low arithmetic complexity. As a validation of our method (implemented into the open-source finite-element micromagnetic package \textsc{TetraX}), we present results for various systems and compare them with theoretical predictions and with established finite-difference methods. We believe this method offers an efficient and versatile tool to calculate spin-wave dispersions in layered magnetic systems.

\end{abstract}

\maketitle

\section{Introduction}



Over the last decades, micromagnetic simulations have become a powerful method to predict new effects in magnetism besides being used to complement experimental results and to validate analytical derivations. The continuous requirement to simulate the static magnetic states and magnetization dynamics in complex systems has lead to the development of highly-optimized micromagnetic codes\cite{donahueOOMMFUserGuide1999,kakaySpeedupFEMMicromagnetic2010,changFastMagFastMicromagnetic2011,vansteenkisteDesignVerificationMuMax32014} for classical time-integration-based methods that solve the equation of motion of the magnetization on a discrete mesh. Certainly, these codes have already played a key role in the investigation of the fundamental as well as applied aspects of magnetism, such as magnetization reversal processes,\cite{schumacherQuasiballisticMagnetizationReversal2003,hertelExchangeExplosionsMagnetization2006} domain wall dynamics,\cite{parkinMagneticDomainWallRacetrack2008,klauiDomainWallMotion2003,yanFastDomainWall2011,otaloraChiralitySwitchingPropagation2012,yanChiralSymmetryBreaking2012} skyrmion motion driven by spin-polarized currents,\cite{fertSkyrmionsTrack2013,buttnerDynamicsInertiaSkyrmionic2015} non-linear magnetization dynamics for reservoir and neuromorphic computing,\cite{huangMagneticSkyrmionbasedSynaptic2017,abeedEffectMaterialDefects2019,songSkyrmionbasedArtificialSynapses2020,korberNonlocalStimulationThreemagnon2020,grollierNeuromorphicSpintronics2020} or the hunt for exotic magnetic textures such as hopfions,\cite{bogolubskyThreedimensionalTopologicalSolitons1988,borisovThreedimensionalStaticVortex2010a,rybakovMagneticHopfionsSolids2019,voinescuHopfSolitonsHelical2020} just to name a few. However, even with such micromagnetic codes, the numerical study of spin-wave propagation in waveguides or extended films remained ineffective due to the required long simulation times and large mesh dimensions for proper frequency and wave-vector resolution. Other pitfalls are the extensive post processing required after each simulation, the necessity to have \textit{a priori} knowledge about the mode-profile symmetries, the inability to detect degenerate modes, and other related problems.\cite{henryPropagatingSpinwaveNormal2016,korberFiniteelementDynamicmatrixApproach2021} This, finally, initiated the development of different approaches. Among those best suited to study linear spin-wave dynamics is the dynamic-matrix method,\cite{giovanniniSpinExcitationsNanometric2004,naletovIdentificationSelectionRules2011,taurelCompleteMappingSpinwave2016,brucknerLargeScaleFiniteElement2019} especially the approach for propagating spin waves in systems with a translationally invariant magnetic equilibrium.\cite{hillebrandsSpinwaveCalculationsMultilayered,henryPropagatingSpinwaveNormal2016}

In contrast to standard time-domain micromagnetic simulations, which rely on the time-integration of the nonlinear Landau-Lifshitz-Gilbert (LLG) equation of motion of the magnetization and subsequent post-processing of the simulation data by means of Fourier analysis to obtain spin-wave frequencies and spatial mode profiles, dynamic-matrix approaches (or dynamic-matrix methods) are based on an exact numerical solution of the linearized LLG equation about some (unitary) equilibrium magnetization $\bm{m}_0(\bm{r})$. The linearized equation of motion for the complex and unitless spatial mode profiles $\delta \bm{m}(\bm{r})$ of the spin-wave modes reads as
\begin{equation}\label{eq:linearized_eq}
  \frac{\mathrm{d}}{\mathrm{d}t}(\delta \bm{m}) = -\omega_M(\delta \bm{h} \cross \bm{m}_0 + \delta \bm{m} \cross \bm{h}_0)
\end{equation}
with $\omega_M = \gamma \mu_0 M_s$ being the characteristic magnetic frequency, $\gamma$ being the modulus of the gyromagnetic ratio and $M_\mathrm{s}$ being the saturation magnetization of the magnetic body at hand, the vectors $\bm{h}_0$ and $\delta \bm{h}$ are the unitless static and dynamic effective fields.

Considering a waveguide with at least one of its dimensions extended to infinity, the linearized equation can be transformed into a plane-wave problem. The spin-wave mode profiles for a magnetic waveguide with translationally invariant equilibrium can be written as
\begin{equation}\label{eq:mode_profile}
  \delta\bm{m} \propto \bm{m}_k \equiv  \bm{\eta}_k(x,y)e^{i(kz-\omega t)},
\end{equation}
with $k$ being the wave number in the direction of propagation ($z$ in this manuscript) and $\bm{\eta}_k$ being the complex lateral mode profile. Using this definition and after some transformations (see \textit{e.g.} Ref.~\citenum{korberFiniteelementDynamicmatrixApproach2021}), the linearized equation Eq.~\eqref{eq:linearized_eq} becomes a wave-vector-dependent eigenvalue problem
\begin{equation}\label{eq:eigenvalue-problem}
  \frac{\omega(k)}{\omega_M} \bm{\eta}_k = \vu{D}_k \bm{\eta}_k
\end{equation}
that yields the frequencies of the eigenmodes $\omega(k)$ and their lateral spatial profiles $\bm{\eta}_k$ simultaneously for wavelengths far greater than the lattice parameter of the investigated magnetic specimen. Here, $\vu{D}_k$ is the so-called dynamic matrix which, in the case of propagating waves, depends on the wave vector $k$. After spatial discretization, the dynamic matrix $\vu{D}_k$ can be diagonalized for each $k$ using a suitable numerical eigensolver. Apart from arbitrary frequency- and wave-vector resolution, a major benefit in adapting the numerical modeling to plane waves is that only the lateral spatial directions $(x,y)$ need to be discretized. This means that, for example, in the case of a waveguide, only a single two-dimensional cross section needs to be modeled. In the case of infinite film, even only a line-trace along its normal direction needs to be modeled.\cite{hillebrandsSpinwaveCalculationsMultilayered,henryPropagatingSpinwaveNormal2016,gallardoSpinwaveNonreciprocityMagnetizationgraded2019}

Recently, we presented a finite-element dynamic-matrix approach that allows to calculate the spin-wave dispersion and related mode profiles in translationally invariant waveguides with arbitrary cross section.\cite{korberFiniteelementDynamicmatrixApproach2021} To take into account the (usually) computationally demanding dipolar fields, we presented an extension of the finite-element/boundary-element method by Fredkin and Koehler\cite{fredkinHybridMethodComputing1990} to calculate the lateral dipolar potentials of propagating spin waves. This method relies on a numerical solution of the Poisson equation, which governs the dipolar potential of each mode, and comes with the major advantage that only the magnetic material needs to be modeled. Moreover, it keeps the resulting eigenvalue problem Eq.~\eqref{eq:eigenvalue-problem} sparse, leading to a lower arithmetic complexity than explicitly calculating the dipolar tensor in matrix form, which is usally done in finite-difference codes. Furthermore, discretizing the cross section of a waveguide using triangular finite elements allowed to model the spin-wave dynamics in waveguides of any cross-section shape, such as polygonal tubes,\cite{korberSymmetryCurvatureEffects2021a,korberModeSplittingSpin2022a} thick-shell round nanotubes,\cite{korberCurvilinearSpinwaveDynamics2022} rectangular waveguides\cite{korberFiniteelementDynamicmatrixApproach2021, korberNumericalReverseEngineering2021} and many more, as long as the area of the cross section remained finite.

Naturally, the important cases of mono- and multilayer systems cannot be captured by this method. Therefore, in this paper, we extend our finite-element dynamic-matrix approach for propagating waves to infinitely extended mono- or multilayers, which, notably, can have arbitrary thicknesses and spacings between each other. We also note that the equilibrium magnetization and spin-wave dynamics can be inhomogeneous along the thickness of each individual layer. For the extension, we derive analytically the most critical part, namely the computation of the dipolar fields, by extending the plane-wave Frekdin-Koehler method to infinite layers. 
As this method avoids the discretization of the non-magnetic material, the computational effort is completely independent of the spacer thickness, which solely appears as a numerical factor in the computations. To validate our method, we first compare our results for the dipole-exchange spectra of spin waves in monolayers with varying film thickness with well-known analytical formulae from the literature, as well as with established finite-difference propagating-wave dynamic-matrix approaches.\cite{henryPropagatingSpinwaveNormal2016,gallardoDipolarInteractionInduced2018} Furthermore, we
validate our method for magnetic bilayer systems, when the ferromagnetic layers are separated by a non-magnetic spacer with varying spacing distance, by comparing the dispersion of the symmetric and antisymmetric modes to the analytical derivations from \textcite{gallardoReconfigurableSpinWaveNonreciprocity2019} As an important contribution for the study of multilayer systems, we also introduce interlayer-exchange interaction into our FEM dynamic-matrix approach. The presented method is readily implemented in the \textsc{Tetrax} open-source micromagnetic modeling package.\cite{korberTetraXFiniteElementMicromagneticModeling2022}


\section{Numerical implementation of effective fields}

The construction of the dynamic matrix from Eq.~\eqref{eq:eigenvalue-problem} requires the evaluation of the magnetic interactions and their related stiffness fields as in usual micromagnetic simulations. Therefore one may need to take into account contributions of the various interactions, such as exchange (being symmetric or asymmetric), dipole-dipole, magnetocrystalline, interlayer-exchange and Zeeman interaction. Here, we give a detailed consideration of the dipolar field, as being the main focus of the paper, while also introducing expressions for the symmetric (intralayer) exchange as well as the interlayer-exchange interaction.

\subsection{Dipolar field}

\subsubsection{Screened Poisson equation of propagating spin waves}
\label{subsec:FredkinKoehler}

Taking into account the effect of dynamic dipolar fields on spin-wave dispersions requires calculating the dipolar field (or demagnetizing field) of each mode as
\begin{equation}
  \bm{h}_k^{({\mathrm{dip})}} = -\nabla \phi_k \equiv -\vu{N}_k^{({\mathrm{dip})}}\bm{m}_k,
\end{equation}
with $\phi_k(\bm{r}, t)$ being the unitless dipolar potential (or magnetostatic potential) generated by each mode. The linear operator $\vu{N}_k^{({\mathrm{dip})}}$ which outputs the dipolar field at wave-vector $k$ is also referred to as the plane-wave dipolar (or demagnetizing) tensor.\cite{henryPropagatingSpinwaveNormal2016} Considering some magnetic waveguide which is translationally invariant along the $z$ direction [see Fig.~\figref{fig:FIG1}{a}], the potential of spin waves propagating along this direction with wave number $k$ is of the form
\begin{equation}
  \phi_{k}(\bm{r}, t) = \psi_{k} (\bm{\rho}) e^{i[kz - \omega(k)t]}.
\end{equation}
Here, $\psi_k(\bm{\rho})$ is the complex lateral potential of the respective mode, which is defined only within the cross section $A$ of the magnetic element, which, in return, is some arbitrarily shaped subset of the $(x,y)$ plane [see Fig.~\figref{fig:FIG1}{a}]. The lateral potential $\psi_{k} (\bm{\rho})$ can be obtained by solving the screened Poisson equation
\begin{subequations}\label{eq:screenedPoissoneq}
\begin{equation}
  (\nabla^2 - k^2) \psi_{k} = \left\{\begin{array}{ll} (\nabla+ ik\vu{e}_z)\bm{\eta}_{k} & \text{for}\ \bm{\rho} \in A, \\
       0 & \text{elsewhere}\end{array}\right. ,
\end{equation}
with the following continuity and jump conditions at the boundary of the magnetic element:
\begin{equation}
\label{eq:ContinuityCondition}
  \psi_\mathrm{out}(\bm{\rho})\vert_{\partial A} - \psi_\mathrm{in}(\bm{\rho})\vert_{\partial A} = 0
\end{equation}
and
\begin{equation}
\label{eq:JumpCondition}
  \frac{\partial \psi_\mathrm{out}(\bm{\rho})}{\partial \bm{n}(\bm{\rho})}\bigg \vert_{\partial A} -\frac{\partial \psi_\mathrm{in}(\bm{\rho})}{\partial \bm{n}(\bm{\rho})}\bigg \vert_{\partial A} = - \bm{n}(\bm{\rho}) \cdot \bm{\eta}_{k} (\bm{\rho})\vert_{\partial A},
\end{equation}
\end{subequations}
where $\psi_\mathrm{in/out}$ is the potential inside/outside of the magnetic material and $\bm{n}$ is the normal vector. It is also required that $\psi(\bm{\rho}) \rightarrow 0$ as $\bm{\rho} \rightarrow \infty$.

\begin{figure}
  \centering
  \includegraphics{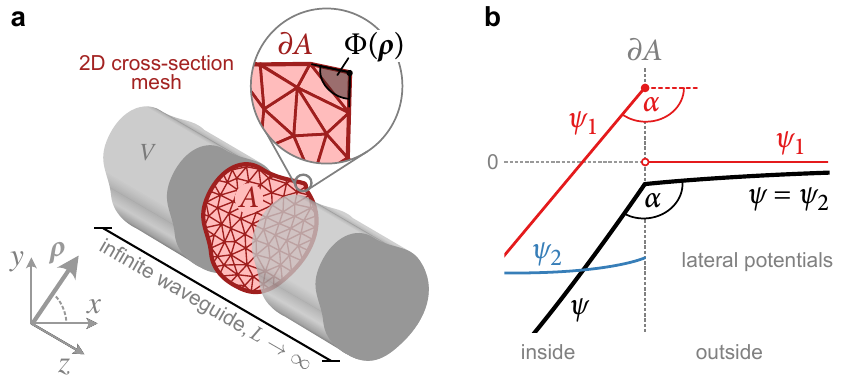}
  \caption{(a) Schematics of an infinitely extended waveguide with arbitrary cross section A. The magnified inset shows the angle attributed to a certain boundary node. (b) The real parts of the different lateral potentials in the plane-wave Fredkin-Koehler method are shown crossing the boundary $\partial A$ of the magnetic sample (Figure adapted from Ref.~\citenum{korberFiniteelementDynamicmatrixApproach2021}).}
  \label{fig:FIG1}
\end{figure}

\subsubsection{Recap: Fredkin-Koehler method for propagating spin waves}

There are several ways to numerically obtain a solution to the boundary-value problem Eqs.~\eqref{eq:screenedPoissoneq}. In general, it is possible to calculate the potential by convolution with the kernel
\begin{equation}\label{eq:Green's-function}
  G_k(\bm{\rho},\bm{\rho}^\prime) = -\frac{1}{2\pi}K_0(\abs{k}\abs{\bm{\rho}-\bm{\rho}^\prime})
\end{equation}
which is the Green's function of the Yukawa operator $\Delta - k^2$ in two dimensions, with $K_0$ being the modified Bessel function of second kind and zeroth order. Performing this convolution numerically is non-trivial due to the divergent and asymptotic behavior of the Green's function. These features can lead to a considerable accumulation of numerical noise and, thus, to an imprecise calculation of the dipolar fields. When discretizing a magnetic body using a structured mesh, as commonly done in the finite-difference (FD) method, this integration can indeed be made much more stable by averaging within the regular cells of the mesh and calculating the kernel for each cell. This, in return, allows to explicitly calculate the dipolar tensor $\vu{N}_k^{(\mathrm{dip})}$ for each wave vector $k$ in matrix form. This was done, for example, for plane layer stacks or rectangular waveguides in Ref.~\citenum{henryPropagatingSpinwaveNormal2016}. Note, however, that this makes the eigenvalue problem Eq.~\eqref{eq:eigenvalue-problem} dense due to the long-range character of the dipole-dipole interaction.

When working with unstructured meshes, as done in finite element methods (FEM), this explicit calculation of the cell kernels is not possible. Instead, to circumvent the asymptotic characteristics of the Green's function, in FEM, the dipolar field of each propagating mode can be calculated by first calculating the lateral potential by explicitly solving the screened Poisson equation with according boundary- and jump conditions [Eq.~\eqref{eq:screenedPoissoneq}]. Note that this keeps the eigenvalue problem Eq.~\eqref{eq:eigenvalue-problem} sparse.\footnote{Although the dipolar interaction is long range, the corresponding differential equation is local.} Recall that to solve the boundary-value problem Eq.~\eqref{eq:screenedPoissoneq}, the boundary conditions at infinity need to be specified. In order to avoid having to model a large "airbox" around the magnetic sample, these boundary conditions can be mapped directly onto the sample surface using a hybrid finite element/boundary element (FEM/BEM) method known as the Fredkin-Koehler method.\cite{fredkinHybridMethodComputing1990} A major benefit of this method is that only the magnetic sample itself needs to be modeled. In our recent work,\cite{korberFiniteelementDynamicmatrixApproach2021} this method has been extended to a plane-wave Fredkin-Koehler method capable of calculating the lateral potential of propagating waves by considering only a single (finite) cross section of the waveguide. Before extending this method further from finite cross sections (waveguides) to infinite cross sections (extended layer stacks), for the reader's convenience, it is worth recalling the basic ideas of the plane-wave Fredkin-Koehler method.

To improve the readability, the subscript $k$ is suppressed for the next part of the manuscript. The idea of the Fredkin-Koehler method is to divide the lateral potential into two parts, such that the first potential fulfills the jump condition in Eq.~\eqref{eq:JumpCondition} at the boundary, which is given by the magnetic surface charges. The second potential, coupled to the first one through an appropriate Dirichlet boundary condition, ensures the continuity of the whole potential $\psi = \psi_1 + \psi_2$. The role of these two potentials is sketched in Fig.~\figref{fig:FIG1}{b}. To obtain them, the following set of equations must be solved for the first potential
\begin{subequations}
\begin{align}
      (\Delta - k^2)\psi_1 &  = (\nabla + ik\bm{e}_z)\bm{\eta} \quad
  \text{in} \, A,\label{eq:poisson-eq}\\
      \frac{\partial}{\partial \bm{n}}\psi_1 & = \bm{n} \cdot \bm{\eta} \quad \text{at} \, \partial A,\label{eq:neumann-b}\\
          \psi_1&  = 0 \quad \text{outside} \, A\label{eq:psi-outside}
\end{align}
and for the second potential
\begin{align}
     (\Delta - k^2)\psi_2 &= 0 \quad \text{in} \, A,\\
         \psi_2 & = u(\bm{\rho}) \quad \text{at} \, \partial A,\\
             (\Delta - k^2) \psi_2 &= 0 \quad \text{outside} \, A.
\end{align}
\end{subequations}
where $u(\bm{\rho})$ denotes the Dirichlet boundary condition for $\psi_2$, which is of the form
\begin{align}
\begin{split}
\label{eq:plane_wave_boundary_condition}
  u(\bm{\rho}) = \frac{1}{2\pi} \oint_{\partial A} \, ds' \psi_1(\bm{\rho'}) \frac{\partial}{\partial \bm{n}} K_0(|k| \cdot | \bm{\rho}- \bm{\rho'}|) \\
  + (\frac{\Phi(\bm{\rho})}{2\pi} -1)\psi_1(\bm{\rho}), \quad \bm{\rho} \in \partial A
\end{split}
\end{align}
and can be calculated after $\psi_1$ is computed by solving the sparse linear system corresponding to Eq.~\eqref{eq:poisson-eq}. Here $\Phi(\bm{\rho})$ is the boundary angle subtended by the boundary point $\bm{\rho}$ within a cross section [see zoom-in in Fig.~\figref{fig:FIG1}{a}], and $ds'$ is the line element on $\partial A$. Let us note that Eq.~\eqref{eq:plane_wave_boundary_condition} is the plane-wave version of the boundary condition, which can be derived (as done in Ref.~\citenum{korberFiniteelementDynamicmatrixApproach2021}) directly from the original relation for the regular three-dimensional Poisson problem from Fredkin and Koehler.\cite{fredkinHybridMethodComputing1990} When discretizing the magnetic body into finite elements, all differential operators ($\nabla$, $\Delta$ and so forth) take the form of sparse matrices. Furthermore, the Dirichlet boundary condition Eq.~\eqref{eq:plane_wave_boundary_condition} can be expressed using the Dirichlet matrix $\underline{\vu{B}}$ such that
\begin{equation}
  \underline{\psi_2}= \underline{\Hat{\mathbf{B}}_k}\; \underline{\psi_1}
\end{equation}
with $\underline{\psi_{1,2}}$ being the mesh vectors of the two potentials at the boundary. It is clear, that in order to extend the plane-wave Frekdin-Koehler method to infinitely extended layer stacks, we need to derive a tractable expression for the Dirichlet matrix $\vu{B}_k$ which can be evaluated as the cross section of the waveguide becomes infinite in one direction. Note that, $\underline{\Hat{\mathbf{B}}_k}$ is a dense matrix of size $n_B\times n_B$ (with $n_B$ being the number of boundary nodes). This means that the Frekdin-Koehler method re-introduces a dense-matrix multiplication into the solution of the eigenvalue problem Eq.~\eqref{eq:eigenvalue-problem}. We will see soon, however, that the Dirichlet matrix for extended layer stacks will be very small. For the following discussion it is important to remark that this method can also be used if the magnetic region consists of several disjoint regions.

\subsubsection{Extension to mono- and multilayers}

\begin{figure}
  \centering
  \includegraphics[width=\linewidth]{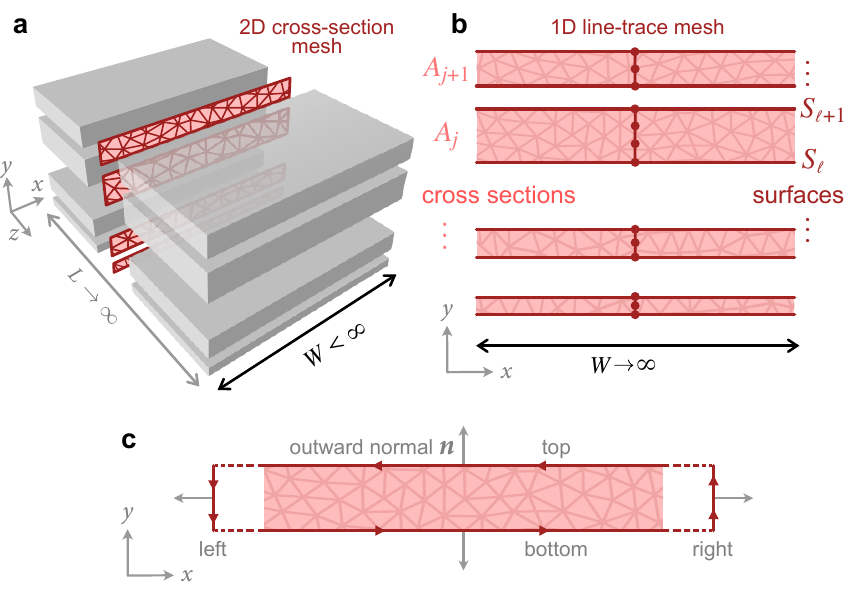}
  \caption{(a) Schematics of a multilayer waveguide with finite width $W$ composed of magnetic layers separated by non-magnetic spacers, both with arbitrary film thicknesses. Extending the lateral dimension to infinity will allow to model the layers using a one-dimensional line-trace mesh along the thickness ($y$ direction) of each individual ferromagnetic layer only, as shown in (b). (c) The boundary integral in Eq.~\ref{eq:green_boundary} can be split along the circumference of each cross section into integrals for the different edges with a fixed integration direction. The normal vector $\bm{n}$ is defined pointing outwards.}
  \label{fig:FIG2}
\end{figure}

To extend the plane-wave Fredkin-Koehler method to mono- and multilayers, we consider a stack of rectangular waveguides with finite width $W$, as exemplified in Fig.~\figref{fig:FIG2}{a}, which we can describe with the aforementioned method. Let $A_j$ be the cross section of the $j$-th rectangular waveguide, such that the full cross section is given by $A = \cup_j A_j$. We now want to calculate the boundary condition for the second potential $\psi_2$ [Eq.~\eqref{eq:plane_wave_boundary_condition}] analytically while letting the width of all layers go to infinity, $W \rightarrow \infty$ along the $x$ direction, as seen in Fig.~\figref{fig:FIG2}{b}. This allows to model the whole layer stack using only a single one-dimensional (1D) line-trace mesh along its thickness ($y$) direction, as also seen in Fig.~\figref{fig:FIG2}{b}. Note that, since the Fredkin-Koehler method avoids modeling non-magnetic regions, the mesh does not have to be connected and the resolution can be freely varied for the different layers.

As a first observation, we see that, as $W\rightarrow \infty$ the first lateral potential $\psi_1$ becomes independent of $x$ and can only depend on the thickness coordinate $y$, thus $\psi_1 = \psi_1(y)$. This is clear from the fact that the lateral mode profiles can only depend on the thickness direction in a thick film [$\bm{\eta}=\bm{\eta}(y)$] since we set the propagation direction in $z$ direction. Therefore, the Poisson-Neumann problem Eqs.~({\ref{eq:poisson-eq}-\ref{eq:psi-outside}) for $\psi_1$ becomes translationally invariant along the $x$ direction.

Let us proceed with the calculation of the Dirichlet boundary condition $u$ for $\psi_2$ from $\psi_1(y)$. We denote the first term of Eq.~\eqref{eq:plane_wave_boundary_condition} as the Green's function contribution $u_G(\bm{\rho})$ and the second term as the boundary-angle contribution $u_\Phi(\bm{\rho})$. The latter is quite trivially $u_\Phi(\bm{\rho})=-\psi_1(y)/2$, since the boundary angle for a rectangular element is always $\Phi(\bm{\rho})=\pi$, except at the corners of the rectangle which move towards infinity, as $W\rightarrow\infty$. Furthermore, the Green's function contribution separates into a sum over the layers $j$. With this, we have
\begin{equation}
\label{eq:full_boundary_condition}
  \psi_2(\bm{\rho}) = -\frac{1}{2}\psi_1(y) +  \sum\limits_{\substack{j \in \, \text{layers}}} u_{G,j}(\bm{\rho})  \quad \text{at} \, \partial A
\end{equation}
as the boundary condition.
What remains is the calculation of the Green's function $u_{G,j}(\bm{\rho})$ contribution of each rectangular waveguide.
\begin{equation}\label{eq:green_boundary}
  u_{G,j}(\bm{\rho}) = \frac{1}{2\pi} \oint\limits_{\partial A_j} \mathrm{d}s' \psi_{1,j}(\bm{\rho'}) \frac{\partial}{\partial \bm{n}} K_0(\abs{k} \cdot \abs{ \bm{\rho}- \bm{\rho'}}).
\end{equation}
We can split the integral along the circumference of each cross section into the different edges as shown in Fig.~\figref{fig:FIG2}{c},
\begin{equation}
  \oint_{\partial A_j} =  \int_{\rightarrow,\mathrm{bottom}} + \int_{\uparrow,\mathrm{right}} + \int_{\leftarrow,\mathrm{top}} +  \int_{\downarrow,\mathrm{left}}
\end{equation}
Since the Green's function Eq.~\eqref{eq:Green's-function} goes to zero as $W \rightarrow \infty$ the integrals over the side facets (left and right) vanish and can, therefore, safely be ignored (see supplementary material for proof). With that, we can change the summation of layers and top- and bottom surfaces to a summation over all remaining surfaces $S_\ell$ of the layer stack [see Fig.~\figref{fig:FIG2}{b}].
\begin{equation}
  \sum\limits_j u_{G,j}^\mathrm{top}(\bm{\rho}) + u_{G,j}^\mathrm{bottom}(\bm{\rho}) \longrightarrow \sum\limits_{\substack{\ell \in \, \text{surfaces}}} u_{G,\ell} (\bm{\rho})
\end{equation}
Executing the normal derivative (using $K_0^\prime = -K_1$) and inserting $\bm{\rho}=(x,y,0)$, we have as the contribution of each remaining surface
\begin{widetext}
\begin{equation}
      u_{G,\ell}(x,y) = \frac{1}{2\pi} \int\limits_{-W/2}^{W/2}  \mathrm{d}x^\prime\, \psi_{1}(y_\ell) \ \frac{n_\ell (y-y_\ell)}{\sqrt{(x-x^\prime)^2 + (y-y_{\ell})^2}}\abs{k} K_1\Big(\abs{k}\sqrt{(x-x^\prime)^2 + (y-y_{\ell})^2}\Big).
\end{equation}
\end{widetext}
Here $n_\ell =\pm 1$ is the sign (up or down) of the outward-normal direction of each surface [see Fig.~\figref{fig:FIG2}{c}]. As we let $W\rightarrow\infty$, the integral over $x^\prime$ becomes translationally invariant, and, therefore, independent of $x$. Therefore, we can safely set $x=0$. With the abbreviation $\Delta y_\ell = y-y_\ell$ the Green's-function contribution $u_{G,\ell}$ takes the form
\begin{equation}
   u_{G,\ell}(y) = \frac{n_j\psi_{1}(y)}{2\pi} \cdot I(k, \Delta y_\ell)
\end{equation}
with
\begin{equation}
\begin{split}
  I(k, \Delta y_\ell) = \int\limits_{-\infty}^\infty \mathrm{d}x^\prime\,    &\frac{\abs{k}\Delta y_\ell}{\sqrt{(x^\prime)^2 + (\Delta y_\ell)^2}}\\ &\times K_1\Big(\abs{k}\sqrt{(x^\prime)^2 + (\Delta y_\ell)^2}\Big).
\end{split}
\end{equation}
As carried out in the supplementary material, this integral can be solved in a closed form and one arrives at the tractable expression
\begin{equation}\label{eq:integral-closed-form}
\begin{split}
  I(k, \Delta y_\ell) = \mathrm{sgn}(\Delta y_\ell)\,\pi\, e^{-\abs{k}\abs{\Delta y_\ell}}
\end{split}
\end{equation}
with "sgn" denoting the sign function, using the convention $\mathrm{sgn}(0) =0$.
As a result, the whole Dirichlet boundary condition takes the very simple form
\begin{equation}
\begin{split}
      u(y) = & -\frac{\psi_1(y)}{2}\\ &+\sum_{\substack{\ell \in \, \text{surfaces}}} \frac{n_\ell\mathrm{sgn}(y-y_\ell)}{2} \psi_{1}(y_\ell) \  e^{-\abs{k}\abs{y-y_\ell}}
\end{split}
\end{equation}
with, again, $\ell$ running over the boundary surfaces (boundary nodes of the 1D mesh) and $n_\ell$ being the normal direction along the $y$ axis of each surface. It is important to note that for infinite layers (1D samples), the Dirichlet boundary conditions are calculated without having to perform any numerical integration, like is the case for waveguides with finite (2D) cross section or volumetric (3D) samples.

Furthermore, we see that, for infinite layers, the boundary matrix $\vu{B}$ will always be only of size $2N\times 2N$ (with $N$ being the number of layers) and, therefore, its size will be completely independent of the actual thickness of the layers. This fact has considerable implications on the arithmetic complexity of the plane-wave Frekdin-Koehler method for infinite layers. Take, for example, the case of a single monolayer, where $\vu{B}$ contains only four elements. With increasing number of nodes $n$ along the thickness of the layer, only the discretized differential operators ($\Delta$, $\nabla$ etc.), which are all sparse, increase in size. With that, on a 1D mesh, the number of non-zero elements in these sparse matrices only increases as $\mathcal{O}(n)$. In contrast to this, explicitly calculating the matrix elements of the dipolar tensor $\vu{N}_k^\mathrm{(dip)}$ for the same layer and number of cells along the thickness leads to a dense matrix with the number of non-zero elements scaling as $\mathcal{O}(n^2)$.

\subsection{Interlayer-exchange field}

\begin{figure}[h!]
  \centering
  \includegraphics{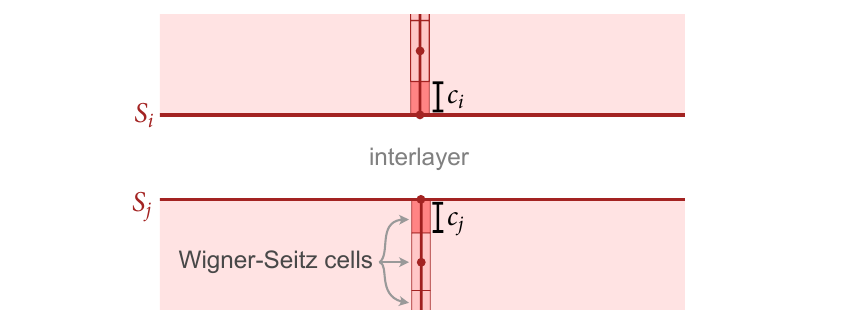}
  \caption{Schematics of two ferromagnetic layers separated by a non-magnetic interlayer and exchange coupled due to RKKY interaction. $S_i$ and $S_j$ are the surfaces of the bilayer system adjacent to the interlayer, while $c_i$ and $c_j$ are the "volumes" (being the length in the case of the 1D line-trace mesh) of the Wigner-Seitz cells associated to the boundary nodes.}
  \label{fig:FIG3}
\end{figure}

To study spin-wave dynamics in multilayer systems it is often also desirable to model a possible interlayer exchange coupling between the different layers.
If two ferromagnetic layers are separated by a metallic non-magnetic interlayer, the exchange coupling due to Ruderman-Kittel-Kasuya-Yosida~\cite{rudermanIndirectExchangeCoupling1954,kasuyaTheoryMetallicFerro1956,yosidaMagneticPropertiesCuMn1957} (RKKY) interaction between them gives rise to the following bilinear energy
\begin{equation}\label{eq:iec-energy}
  E_\mathrm{IEC} = -J_\mathrm{bl}\int\limits_\Gamma \mathrm{d}s\ \bm{m}(\bm{r})\cdot\bm{m}\big(P(\bm{r})\big)
\end{equation}
where $J_\mathrm{bl}$ (given in \si{\joule/\square\meter}) is the bilinear interlayer-exchange constant, $\Gamma$ is the surface of the bilayer system adjacent to the interlayer and $P(\bm{r})$ maps the point $\bm{r}$ on the surface of one magnetic layer to the nearest point on the surface of the other magnetic layer.\cite{abertMicromagneticsSpintronicsModels2019} Equation~\eqref{eq:iec-energy} can be obtained by replacing the nearest-neighbor sum in a Heisenberg Hamiltonian with an integral. Variation of the energy according to Eq.~\eqref{eq:iec-energy} leads to the unitless interlayer-exchange field acting only on the surface $\Gamma$ as
\begin{equation}
  \begin{split}
      \bm{h}^\mathrm{(exc)}(\bm{r}) & = \frac{J_\mathrm{bl}}{\mu_0 M_\mathrm{s}^2} \bm{m}\big(P(\bm{r})\big) \quad \text{only at}\ \Gamma\\
      & \equiv -\vu{N}^\mathrm{(IEC)}\bm{m}
  \end{split}
\end{equation}
On a 1D line-trace mesh and using finite-element discretization, this field can be obtained as a matrix-vector multiplication using the off-diagonal sparse block matrix $\vu{N}^\mathrm{(IEC)}$ with the blocks
\begin{equation}
  \vu{N}^\mathrm{(IEC)}_{ij} = \vu{N}^\mathrm{(IEC)}_{ji} = -\frac{J_\mathrm{bl}}{\mu_0 M_\mathrm{s}^2} \frac{2}{c_i + c_j}\vu{I} \qquad i \neq j
\end{equation}
with $\vu{I}$ being the identity on $\mathbb{R}^3$, $i$ and $j$ being the indices of two coupled boundary elements. Here, on a line-trace mesh, $c_i$ and $c_j$ are the lengths of the Wigner-Seitz cell associated with each boundary node [see Fig.~\ref{fig:FIG3}]. We note that biquadratic interlayer exchange can be considered in a similar way, by including biquadratic terms in the energy functional in Eq.~\eqref{eq:iec-energy}, calculating the corresponding field and subsequently linearizing it with respect to some magnetic equilibrium state.

\subsection{Symmetric-exchange field and other effective fields}

Before being able to calculate the dispersion of spin waves in infinite layers, we also need to consider the (symmetric) internal exchange field of the individual layers. Starting, for example, from the exchange operator for propagating waves in Ref.~\citenum{korberFiniteelementDynamicmatrixApproach2021} and considering the fact that the lateral mode profiles $\bm{\eta}_k$ in infinitely extended layers can only depend on its thickness ($y$) coordinate, the unitless lateral exchange field can be obtained as
\begin{equation}
  \bm{h}_k^\mathrm{(exc)}(y) = \lambda^2_\mathrm{ex} k^2 \bm{\eta}_k(y) - \lambda^2_\mathrm{ex}\frac{\mathrm{d}^2}{\mathrm{d}y^2}\eta_{k,y} (y) \cdot \bm{e}_y
\end{equation}
with $\lambda_\mathrm{ex} = \sqrt{2A_\mathrm{ex} / \mu_0 M_s^2}$ being the exchange length and $A_\mathrm{ex}$ being the exchange stiffness constant of the material. In order to assemble the dynamic matrix $\vu{D}_k$, the differential operator $\mathrm{d}^2/\mathrm{d}^2y$ needs to be discretized on the 1D line-trace mesh of the layer [see again Fig.~\figref{fig:FIG2}{b}] using finite elements under consideration of the exchange boundary condition
\begin{equation}
  \frac{\mathrm{d}}{\mathrm{d}y}\bm{\eta}_k = 0 \quad \text{at}\quad \partial A.
\end{equation}
We note that, using finite elements, the second $y$ derivative on the considered line-trace mesh will be quite similar to the finite-difference version obtained by taking central derivatives.

Deriving expressions for other magnetic interactions such as asymmetric exchange interaction (of bulk- or interface origin), or magneto-crystalline anisotropies in the case of propagating waves in infinite layers works in an analogous and straightforward way. Therefore, these fields are not presented here. As the main result of this work is the calculation of the dynamic dipolar field in infinite layers using finite elements, it is enough to only consider exchange-, dipolar and interlayer-exchange interaction in the following examples.

\section{Validation and applications}

In the remaining part of this paper, we want to validate our FEM dynamic-matrix approach for extended layers for a number of different examples. For this we implemented the developed numerical scheme into the \textsc{Tetrax} open-source micromagnetic modeling package\cite{korberTetraXFiniteElementMicromagneticModeling2022} and will test it by calculating the spin-wave spectra in different mono- and bilayer systems. For our calculations, we adopt typical material parameters of the soft magnetic alloy Ni$_{80}$Fe$_{20}$ as summarized in Tab.~\ref{tab:matparam}.

\begin{table}[h!]
\caption{\label{tab:matparam}Parameters used for micromagnetic modeling.
}
\begin{ruledtabular}
\begin{tabular}{ll}
      exchange stiffness ($A_\mathrm{ex}$) & \SI{11}{\pico\joule/\meter}\\
      saturation ($M_\mathrm{s}$) & \SI{800}{\kilo\ampere/m}\\
      reduced gyromagnetic ratio ($\gamma/2\pi$) & \SI{28}{\giga\hertz/\tesla}\\
      interlayer exchange ($J_\mathrm{bl}$), only Sec.~\ref{sec:afm-layers} & \SI{-0.3}{\milli\joule/\square\meter}
\end{tabular}
\end{ruledtabular}
\end{table}

\subsection{External-field dependence of uniform modes (ferromagnetic resonance)}

As a first example to validate our method, we consider a magnetic monolayer of thickness $d$. For this very simple case, the Dirichlet matrix in the plane-wave Fredkin-Koehler method has only four entries and is given by
\begin{equation}
  \underline{\vu{B}}^{(\mathrm{mono})} = -\frac{1}{2}\mqty(1&\exp(\abs{k}d) \\ \exp(\abs{k}d)&1).
\end{equation}
First, we only calculate the frequencies and spatial profiles of the spin-wave modes at $k=0$ under an applied static external field parallel to the layer [see inset in Fig.~\figref{fig:FIG4}{a}]. At this wave number, $k=0$, the magnetic precession is homogeneous within the layer plane but can still be inhomogeneous along the layer thickness, forming standing waves along the layer thickness. These modes are typically referred to as ferromagnetic-resonance (FMR) modes or perpendicular-standing spin waves (PSSWs) in common microwave-absorption experiments and can be denoted by an index $n$ counting the number of nodal lines along the thickness [see Fig.~\figref{fig:FIG4}{b}]. Their frequency as a function of applied magnetic field is given exactly by [see for example Eq.~(5.18) in Ref.~\citenum{stancilSpinWavesTheory2009}]
\begin{equation}\label{eq:kittel}
  \frac{\omega_n(k=0)}{\omega_M} = \sqrt{(h_\mathrm{ext}+ \lambda_\mathrm{ex}^2\kappa_n^2)(h_\mathrm{ext}+ \lambda_\mathrm{ex}^2\kappa_n^2+1)}
\end{equation}
with $h_\mathrm{ext}=H_\mathrm{ext}/M_\mathrm{s}$ being the unitless static external field and $\kappa_n=n\pi/d$ being the wave number of the perpendicular-standing waves along the thickness ($y$) direction of the monolayer.

In Fig.~\figref{fig:FIG4}{a}, we show the oscillation frequencies of the different PSSWs $n=0,1,2,3$ in a permalloy monolayer of $d=\SI{150}{\nano\meter}$ thickness, in a field range between 0 and \SI{60}{\milli\tesla}, calculated with our dynamic-matrix approach implemented in \textsc{TetraX} (solid lines), showing a perfect agreement with the theoretical prediction according to Eq.~\eqref{eq:kittel} (dashed lines). Note, that the obtained mode profiles along the thickness, shown in Fig.~\figref{fig:FIG4}{b}, are perfect unpinned sinusoidals. For the highest-order mode, $n=3$, which exhibits the shortest wavelength along the thickness, a slight frequency-mismatch can be observed which originates from an underestimation of the exchange interaction, \textit{i.e.}, from insufficient accuracy when calculating the magnetization derivatives along the thickness of the layer ($y$ direction). This, of course, could be improved simply by decreasing the characteristic length of the mesh. Here, the layer has been modeled on a line mesh with an average spacing of \SI{1}{\nano\meter} between the nodes.

\begin{figure}
  \centering
  \includegraphics{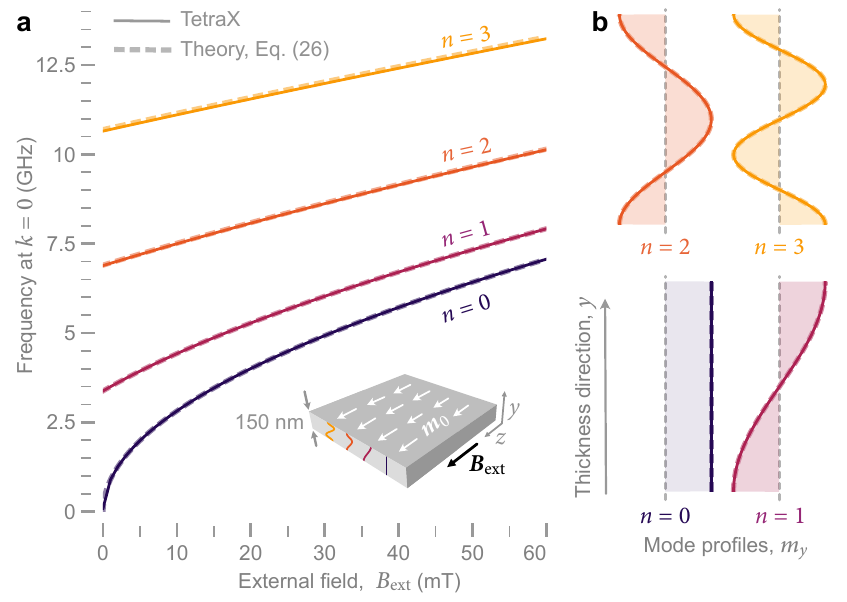}
  \caption{(a) External-field dependence of the uniform mode and the perpendicular-standing waves along the thickness ($n$ being the node number) in a \SI{150}{\nano\meter} thick permalloy film, with the field applied in the plane of the film. A schematics of the magnetic film, the definition of the coordinate system and the equilibrium magnetic state is represented by the inset. In panel (b), the out-of-plane component of the magnetization ($m_y$) along the thickness for the different modes is shown.}
  \label{fig:FIG4}
\end{figure}

\subsection{Spin-wave dispersion in thick films}\label{sec:sw-dispersion-thick-films}

Using the same geometry as in the previous section, we now calculate the dispersion of propagating spin waves ($k\neq 0$), starting with a thin layer of $d=\SI{10}{\nano\meter}$ thickness. Throughout this section, the monolayer is saturated in-plane by a constant external field of $\SI{20}{\milli\tesla}$. As the wave number $k$ departs from zero, the frequency of the spin waves depends crucially on the orientation of their wave vector with respect to the equilibrium magnetization -- a symmetry breaking which is introduced by the dipolar interaction. This can be seen for the two limiting cases of $\bm{k} \parallel\bm{m}_0$ and $\bm{k}\perp\bm{m}_0$ in Fig.~\figref{fig:FIG5}{a}.
Commonly, the spin waves with $\bm{k}\parallel\bm{m}_0$, which propagate parallel to the equilibrium magnetization, are referred to as backward-volume magnetostatic waves (BVMSWs) due to the fact that, with increasing $k$, they are localized mainly to the volume of the layer and can, with increasing layer thickness $d$, exhibit a negative group velocity in certain regions of the wave-vector space. In contrast, the spin waves with $\bm{k}\perp\bm{m}_0$, propagating perpendicular to the equilibrium, generally exhibit a much higher group velocity and, depending on the propagation direction, are localized to either surface of the layer. Hence, they are also referred to as magnetostatic surface waves (MSSWs).

\begin{figure}[h!]
  \centering
  \includegraphics{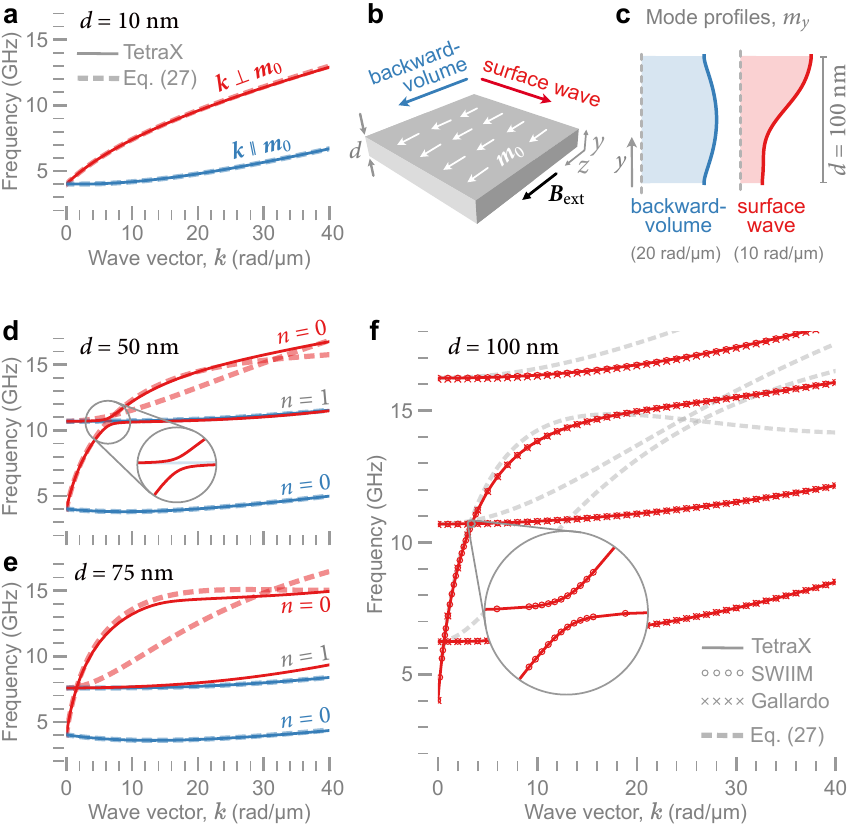}
  \caption{(a) Dispersion relation of the first spin-wave branch in a \SI{10}{\nano\meter} thick film shown for the two main propagation directions. Solid and dashed lines correspond to results of TetraX and predictions of the zeroth-order perturbation theory of Kalinikos and Slavin \cite[Eq.~21]{kalinikosTheoryDipoleexchangeSpin1986}, respectively. (b) Sketch of the film with the used coordinates, the external field direction as well as the two main propagation direction, namely the backward-volume (BV) and surface waves (SW) geometry. Two exemplary mode profiles along the film thickness of the main directions are shown in (c). In (d) and (e), the dispersion relation of a \SI{50}{\nano\meter} and \SI{75}{\nano\meter} thin film is calculated and compared with the analytical predictions (dashed lines). The inset in (d) highlights the dipole-dipole mode hybridization between the first two branches in the SW geometry. In panel (f) the comparison of the dispersion, computed with three different numerical codes, for a \SI{100}{\nano\meter} film shows a perfect agreement between the different numerical codes.  The inset shows that the finite difference code, SWIIM, and our finite element code, TetraX, perfectly overlap even for the computationally most critical branch hybridization region.}
  \label{fig:FIG5}
\end{figure}

From a theoretical point of view, the spin-wave propagation in thin films is most prominently described by the perturbation theory of Kalinikos and Slavin~\cite{kalinikosTheoryDipoleexchangeSpin1986} (KS), in which dipolar fields are calculated in terms of magnetostatic Green's functions. For sufficiently thin layers, neglecting surface pinning and hybridization between different modes, the zeroth-order perturbation of KS provides explicit analytical expressions for the dispersion of the different modes $n$. For the two limiting cases (BVMSW and MSSW), the dispersion can be written as
\begin{subequations}\label{eq:ks}
\begin{equation}
  \frac{\omega_n(k)}{\omega_M} =
  \Big[(h_\mathrm{ext} + \lambda_\mathrm{ex}^2k_n^2) (h_\mathrm{ext}+\lambda_\mathrm{ex}^2k_n^2 + 1 - P_{nn})\Big]
\end{equation}
for $\bm{k}\parallel\bm{m}_0$ (BVMSW) and
\begin{equation}
\begin{split}
     \frac{\omega_n(k)}{\omega_M} = &
  \Big[(h_\mathrm{ext} + \lambda_\mathrm{ex}^2k_n^2+ P_{nn}) \\ & \quad \times (h_\mathrm{ext}+\lambda_\mathrm{ex}^2k_n^2 + 1 - P_{nn})\Big]^{\tfrac{1}{2}}
\end{split}
\end{equation}
for $\bm{k}\perp\bm{m}_0$ (MSSW),
with $k_n^2 = k^2 + \kappa_n^2$ being the square of the \textit{total} wave vector and $P_{nn}$ being given by
\begin{equation}
  P_{nn}=\frac{k^2}{k_n^2}\qty[1-\qty(\frac{2}{1+\delta_{0n}})\frac{k^2}{k_n^2}\qty(\frac{1-(-1)^n e^{-\abs{k}d}}{\abs{k}d})].
\end{equation}
\end{subequations}
For a layer thickness of $d=\SI{10}{\nano\meter}$, we see in Fig.~\figref{fig:FIG5}{a} that our numerical calculations are in perfect agreement with the theoretical prediction made by the zeroth-order perturbation of KS in Eq.~\eqref{eq:ks}. With increasing layer thickness, the higher-order modes ($n>0$) decrease in overall frequency due to decreasing confinement and, therefore, an overall decrease in exchange energy. At the same time, the dispersion curve of the zeroth MSSW mode ($n=0$) acquires a much steeper slope (higher group velocity), as seen for a layer of thickness $d=\SI{10}{\nano\meter}$ in
[Fig.~\figref{fig:FIG5}{d}]. As soon as two modes cross they can, depending on their symmetry, share an avoided level crossing due to dipole-dipole hybridization, an effect, mostly present for the MSSW modes (red lines). For the layer of $d=\SI{50}{\nano\meter}$ thickness [Fig.~\figref{fig:FIG5}{d}], we see how the zeroth $(n=0)$ and the second $(n=1)$ MSSW modes are hybridized, as reflected by an avoided level crossing between them. Naturally, the zeroth-order theory of KS does not capture this feature. Apart from that, already at the thickness of \SI{50}{\nano\meter}, a considerable deviation between our numerical calculations and the analytical theory can be observed, a trend which increases even further with thickness [Figs.~\figref{fig:FIG5}{d-f}]. This, however, is to no surprise, as the zeroth-order KS theory does not consider the perturbation of the spatial mode profiles due to dipolar fields yet. Therefore, it leads erroneous results in thick layers, especially for the MSSW ($\bm{k}\perp\bm{m}_0$) modes as their mode profiles are strongly perturbed by internal dipolar fields. As the zeroth-order theory of KS is still widely used in many works even for thicker samples, it is worth noting that, strictly speaking, this theory is only applicable for very thin layers with thicknesses below the order of a couple of exchange lengths of the respective material (usually below \SI{20}{\nano\meter}). For larger thicknesses, higher-order terms in the perturbation series need to be included, which is already well-described in the seminal work of KS. Recall that a dynamic-matrix approach, such as the one presented here, is a method relying on direct numerical diagonalization of the linearized equation of motion. Therefore, by design, it always provides the exact normal modes of the respective magnetic system (up to discretization errors) and, in principle, is applicable for all thicknesses.

In order to verify the correctness of our finite-element dynamic-matrix approach for thicknesses where Eq.~\eqref{eq:ks} is not valid anymore, we compare our results to calculations performed on the same system using two different finite-difference (FD) codes: \textsc{SWIIM}, developed by Henry \textit{et al.},\cite{henryPropagatingSpinwaveNormal2016} as well as a FD approach by Gallardo \textit{et al.}\cite{gallardoSpinwaveNonreciprocityMagnetizationgraded2019} Both of these methods model the layer on a regular one-dimensional chain of constant spacing along the thickness. Finally, in Fig.~\figref{fig:FIG5}{f}, we show the dispersion of the lowest four MSSW modes ($\bm{k}\perp\bm{m}_0$) calculated for $d=\SI{100}{\nano\meter}$ using \textsc{TetraX} as well as the two FD dynamic-matrix approaches, showing a perfect agreement. For visual clarity, the BVMSWs have been omitted for which the zeroth-order KS theory already provided a good approximation even for larger thicknesses. For completeness, a comparison between \textsc{TetraX} and the two FD codes for all thicknesses between 10 and \SI{75}{\nano\meter} is found in the supplementary material, also showing no discrepancies.

As another suitable display of how accurate dipolar fields are calculated in our method, as an inset in Fig.~\figref{fig:FIG5}{f}, we show a zoom-in on the extremely narrow avoided level crossing which appears due to dipole-dipole hybridization between the zeroth and the second MSSW mode. Here, we also obtain a perfect agreement with the finite-difference calculations. In conclusion, we have verified the correct calculation of dipolar and exchange fields in our numerical scheme of propagating spin waves in a monolayer.

\subsection{Asymmetric spin-wave dispersion in antiferromagnetically-coupled bilayers}\label{sec:afm-layers}

After we have validated the correctness of our method for a single magnetic monolayer, as a final example, we want to extend our consideration to a magnetic bilayer system. In particular, we consider two layers of the same thickness $d$, separated by a non-magnetic spacer of thickness $s$, as depicted in Fig.~\figref{fig:FIG6}{a}. The material parameters are the same as in the previous sections. For such a symmetric layer stack, the $4\times 4$ Dirichlet matrix is given as
\begin{widetext}
\begin{equation}\label{eq:dirichlet-bilayer}
  \underline{\vu{B}}^{(\mathrm{sym\text{-}bi})} = -\frac{1}{2}\mqty(1&\exp(\abs{k}d)&  \exp(\abs{k}(d+s)) & \exp(\abs{k}(2d+s)) \\ \exp(\abs{k}d)&1 & -\exp(\abs{k}s)& -\exp(\abs{k}(d+s)) \\ -\exp(\abs{k}(d+s)) &-\exp(\abs{k}s) & 1&\exp(\abs{k}d) \\ \exp(\abs{k}(2d+s)) & \exp(\abs{k}(d+s)) & \exp( \abs{k}d)&1).
\end{equation}
\end{widetext}
When two magnetic layers are brought into proximity, their spin-wave spectra hybridize (via dynamic dipolar fields or possible dynamic interlayer-exchange fields). This leads, for example, to the mixing of the lowest modes of each layer either into an in-phase (acoustic) or an out-of-phase (optical) mode shown in Fig.~\figref{fig:FIG6}{b}. At $k=0$, that is, for magnetic oscillations homogeneous within the bilayer plane, the optical and acoustic modes are degenerate as long as the layers are not coupled via interlayer exchange, $J_{\mathrm{bl}}= 0$. This degeneracy at $k=0$ is lifted by a non-zero interlayer-exchange coupling, $J_{\mathrm{bl}}\neq 0$, as seen in Fig.~\figref{fig:FIG6}{c}

In case of two layers magnetized antiparallel to each other, spin waves propagating perpendicular to the two magnetizations are non-reciprocal [see Fig.~\figref{fig:FIG6}{a}]. Such a state can be stabilized by an interlayer-exchange coupling with negative sign, $J_\mathrm{bl}<0$, (antiferromagnetic coupling) which, in our case, we set to $J_\mathrm{bl}=\SI{-0.3}{\milli\joule/\square\meter}$.

As a consequence of the antiparallel layer alignment, counter-propagating waves with the same wavelength exhibit different frequencies, seen in Fig.~\figref{fig:FIG6}{c}. This nonreciprocity is purely of dipolar origin and a consequence of magnetochiral symmetry breaking in the pseudo charges generated by the dynamic magnetization. Next to antiparallel alignment of the magnetic layers, this dipolar symmetry breaking can also be introduced by surface curvature,\cite{otaloraCurvatureInducedAsymmetricSpinWave2016,gallardoHighSpinwaveAsymmetry2022,korberCurvilinearSpinwaveDynamics2022} as observed in magnetic nanotubes, or by a chiral magnetic texture, as observed for the spin waves propagating along Bloch walls\cite{henryUnidirectionalSpinwaveChanneling2019} in systems with perpendicular magnetic anisotropy. For the layer stack considered here, the spin-wave spectrum was studied theoretically and experimentally by Gallardo \textit{et al.} in Refs.~\citenum{gallardoReconfigurableSpinWaveNonreciprocity2019,gallardoSpinwaveFocusingInduced2021}. 
%


%
%
%
%
%
%
Finally, in Fig.~\figref{fig:FIG6}{c}, we compare the theoretical dispersion of the acoustic and optical mode in the considered system according to Gallardo \textit{et al.} with the numerical calculations using our FEM dynamic-matrix approach implemented in \textsc{TetraX}, for a layer stack with layer thickness $d=\SI{2}{\nano\meter}$ and spacing $s=\SI{2}{\nano\meter}$. It is possible to see that our numerical scheme is in perfect agreement with the analytical theory of Ref.~\citenum{gallardoReconfigurableSpinWaveNonreciprocity2019} in the case of thin layers. However, analogous to the previous section, as we increase the layer thickness $d$ to $\SI{20}{\nano\meter}$, in Fig.~\figref{fig:FIG6}{d} we can see clear deviations between the theory and our numerical calculations. This again is due to the fact that the theory does not consider dipolar perturbations of the mode profiles, which, for large $d$, become inhomogeneous along the layer thickness. Because of this, in addition, we compare our results again with the FD difference dynamic-matrix approach by Gallardo \textit{et al.},\cite{gallardoSpinwaveFocusingInduced2021} which is also capable of modeling inhomogeneities along the layer thickness. As can be seen in Fig.~\figref{fig:FIG6}{d}, the correctness of our calculations with \textsc{TetraX} is perfectly supported by the FD calculations.

\begin{figure}[h!]
  \centering
  \includegraphics{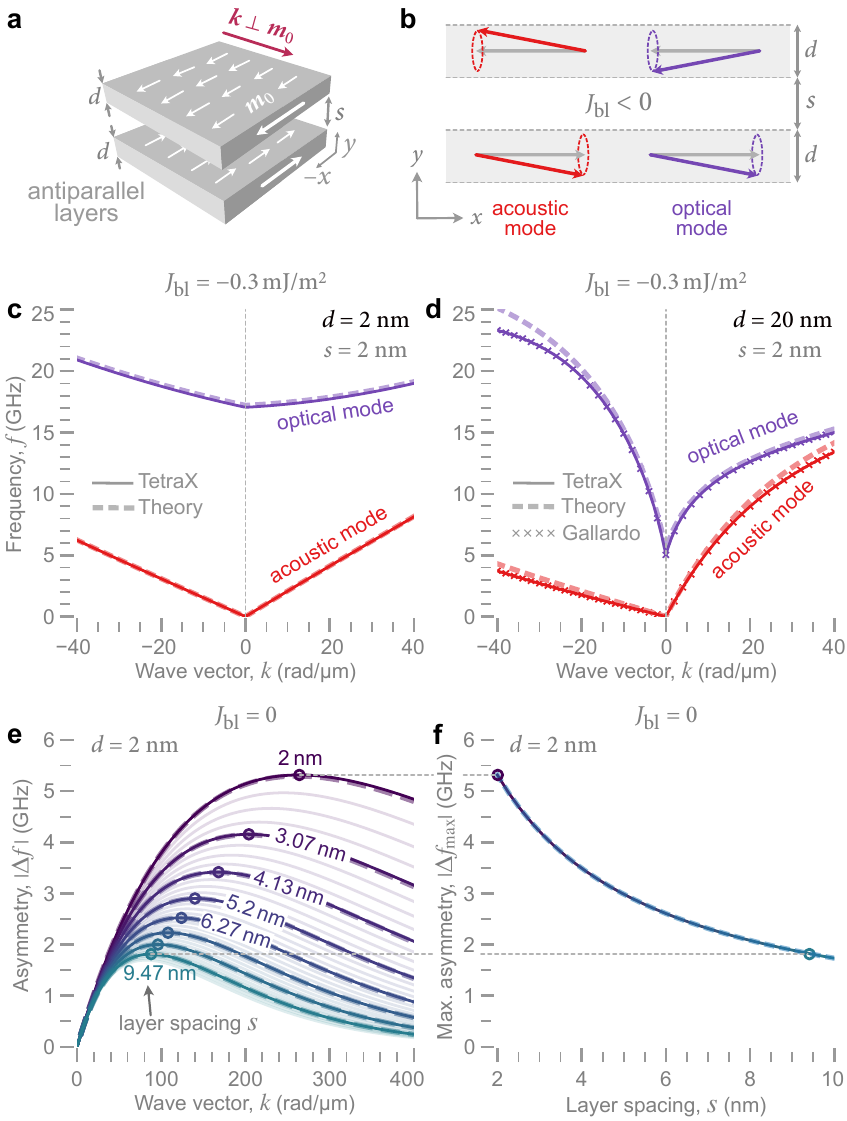}
  \caption{(a) Schematics of a symmetric  bilayer with thickness $d$ and spacing $s$ magnetized antiparallel to each other. The dispersion is calculated for spin waves propagating perpendicular to the static magnetization. The lowest two spin-wave branches are the homogeneous symmetric and antisymmetric modes, with mode profiles sketched in (b). In (c) and (d) the simulated dispersion relation of the modes in (b) in comparison with the analytical solution is shown for bilayers with \SI{2}{\nano\meter} as well as \SI{20}{\nano\meter} thickness and \SI{2}{\nano\meter} separation. As expected and in accordance with previous micromagnetic simulations, for the larger film thickness the analytical results deviate from the simulation results using \textsc{TetraX}. Instead, in (d), we compare our results also to the finite-difference dynamic-matrix results (crosses) of Gallardo \textit{et al.} according to Ref.~\citenum{gallardoSpinwaveFocusingInduced2021}. The spin-wave asymmetry versus the propagation vector for a variety of interlayer spacers is summarized in panel (e). Finally, in (f), the maximum asymmetry as a function of the layer spacing is shown for layers with \SI{2}{\nano\meter} thickness. For all panels, the dashed lines are the analytical predictions.}
  \label{fig:FIG6}
\end{figure}

Let us highlight here a major benefit of the plane-wave Fredkin-Koehler method that we use to calculate the dynamic dipolar fields: As this method avoids for the non-magnetic material (here: the spacer/interlayer) to be part of the mesh, the computational effort is completely independent of the exact value of the spacer thickness $s$ which solely appears as a parameter in the Dirichlet matrix Eq.~\eqref{eq:dirichlet-bilayer}. Not only is it easy to continuously vary the spacing $s$. It is also possible to take the limit of extremely large or small $s$ without ever increasing the number of nodes in the mesh and, therefore, without increasing the computational effort. \textcolor{black}{We note that the same is true when explicitly calculating the plane-wave dipolar tensors $\vu{N}^\mathrm{(dip)}_k$ in matrix form, as done, for example, in the FD code SWIIM.\cite{henryPropagatingSpinwaveNormal2016} In this case, too, only the magnetic material needs to be modeled (for the convolution with the Green's function), while any inter-cell spacings appear as parameters in the dipolar tensors. Recall, however, that this makes the dynamic-matrix and, therefore, the eigenvalue problem, dense.}

\textcolor{black}{To illustrate the flexibility with respect to changing the interlayer spacings in our FEM dynamic-matrix approach}, we present how the dipole-induced dispersion asymmetry in the considered layer stack changes when varying the spacing $s$, for a fixed $d=\SI{2}{\nano\meter}$. Technically, the sign and magnitude of the interlayer-exchange coupling will of course vary with the thickness $s$ of the interlayer. Therefore, we will completely disregard it here, $J_\mathrm{bl}=0$. This is reasonable in the sense that, for thin layers, interlayer-exchange coupling has no direct quantitative influence on the dispersion asymmetry.

In Fig.~\figref{fig:FIG6}{e}, we show the asymmetry $\abs{\Delta f} = \abs{f(k) - f(-k)}$, which, for the case of thin layers, has the same magnitude for the acoustic and optical mode. We start from $s=\SI{2}{\nano\meter}$, which corresponds to the same case as shown in Fig.~\figref{fig:FIG5}{c}, and go up to $s=\SI{10}{\nano\meter}$. All solid curves have been calculated using \textsc{TetraX} in approximately the same amount of time (less than a minute). For visual clarity, the analytical results according to Ref.~\citenum{gallardoReconfigurableSpinWaveNonreciprocity2019} are only shown for selected spacings. It is possible to see that, with increasing spacing $s$, the position of the maximum dispersion asymmetry shifts to lower wave numbers $\abs{k}$ while its maximum value decreases. As a figure of merit, in Fig.~\figref{fig:FIG6}{f} we show the smooth transition of this maximum as a function of spacing $s$, again, showing perfect agreement between numerics and analytics.

\section{Conclusions}

In summary, we have extended our finite-element dynamic-matrix approach for propagating spin waves for waveguides of arbitrary cross section to mono- and multilayers of arbitrary spacing and thickness. Therefore the dispersion relation for extended films can be calculated by using an 1D line-trace mesh only along the thickness of the ferromagnetic film. To do so, the previously presented Fredkin-Koehler method (also known as the hybrid finite-element / boundary element method) to solve the screened Poisson equation of propagating spin waves was extended for mono- and multilayers, allowing to compute the dipolar potential and related stiffness field in a very efficient manner. Remarkably, the obtained boundary matrix (or Dirichlet matrix) has exact elements, defined by analytical expressions. The major benefit of this method is that it avoids the discretization of any non-magnetic material while also keeping the resulting eigenvalue problem sparse. In particular, this means that the computational effort is completely independent on the spacer thickness, which solely appears as a parameter in the Dirichlet matrix. Moreover, the resulting matrices only scale with the number of nodes $n$ along the normal direction of the layers as $\mathcal{O}(n)$, whereas it scales as $\mathcal{O}(n^2)$ when explicitly calculating the dipolar tensors. This, provides our FEM approach with a comparably low arithmetic complexity.

Our method has been validated for a number of known systems using theoretical predictions as well as finite-difference implementations established previously. The presented method is readily implemented into the \textsc{TetraX}~\cite{korberTetraXFiniteElementMicromagneticModeling2022} open-source micromagnetic modeling package offering for the magnonic community an easy and efficient calculation of spin-wave dispersions for various standard magnonic problems.

\section*{Supplementary material}

See the supplementary material, which includes Ref.~\citenum{NIST:DLMF}, for the derivation of the closed solution of the integral $I(k,\Delta y_\ell)$ in  Eq.~\eqref{eq:integral-closed-form} and an extended comparison between the numerical results of the finite-difference solvers SWIIM and the one by Gallardo \textit{et al.} with our approach finite-element approach for the examples discussed in Sec.~\ref{sec:sw-dispersion-thick-films}.

\section*{Author's contributions}

L.K and A.H. contributed equally to this work. L.K. and A.K. conceptualized this work. L.K., A.H. and A.O. derived the analytical expressions for plane-wave Fredkin-Koehler method in multilayers.  A.H, L.K. and A.K. implemented the numerical code.
L.K., A.H. and A.K. co-wrote the manuscript. R.A.G. and Y.H. calculated reference data for the validation of the method. All authors discussed and revised the final manuscript.

\section*{Acknowledgements}

The authors acknowledge fruitful discussions with Claas Abert on the interlayer-exchange coupling within the finite-element method.
Financial support by the Deutsche Forschungsgemeinschaft within the programs KA 5069/1-1 and KA 5069/3-1 is gratefully acknowledged. RAG acknowledges financial support from Fondecyt, Grants No. 1210607.

\section*{Data availability}
The data that support the findings of this study are openly available in RODARE at \href{TODO}, reference \citenum{}.

%



\end{document}